\documentclass[aps,prl,twocolumn,superscriptaddress]{revtex4}

\usepackage{graphicx}

\newcommand{\si}{\sigma}
\newcommand{\al}{\alpha}
\newcommand{\tet}{\theta}
\newcommand{\az}{\varphi}

\newcommand{\oeq}{\begin{equation}}
\newcommand{\ceq}{\end{equation}}
\newcommand{\oeqn}{\begin{eqnarray}}
\newcommand{\ceqn}{\end{eqnarray}}

\renewcommand{\>}{\rangle}
\newcommand{\<}{\langle}
\renewcommand{\(}{\left(}
\renewcommand{\)}{\right)}

\newcommand{\lll}{\left|}
\newcommand{\rll}{\right|}

\newcommand{\stf}{\,\,\,}
\newcommand{\sdf}{\,\,}
\newcommand{\stb}{\!\!\!}
\newcommand{\sdb}{\!\!}

\newcommand{\kfi}{|\phi \>}

\newcommand{\bfi}{\<\phi |}

\newcommand{\oP}{\hat{P}}

\newcommand{\oad}{\hat{a}^\dagger}
\newcommand{\oa}{\hat{a}}

\newcommand{\oN}{\hat{N}}

\newcommand{\del}{\delta\!}

\renewcommand{\d}{{\mbox d}}

\renewcommand{\vr}{{\bf r}}

\begin{document}

\title{Particle transfer reactions with the time-dependent Hartree-Fock theory using a particle number projection technique}

\author{C\'{e}dric Simenel}\email[]{cedric.simenel@cea.fr}
\affiliation{Department of Nuclear Physics, Research School of Physics and Engineering, Australian National University, Canberra, Australian Capital Territory 0200, Australia}\affiliation{CEA, Centre de Saclay, IRFU/Service de Physique Nucl\'eaire, F-91191 Gif-sur-Yvette, France.}

\date{\today}

\begin{abstract}
A particle-number projection technique is used to calculate transfer probabilities 
in the $^{16}$O+$^{208}$Pb reaction below the fusion barrier.
The time evolution of the many-body wave function is obtained with 
the time-dependent Hartree-Fock (TDHF) mean-field theory.
The agreement with experimental data for the sum of the proton-transfer channels is good, 
considering that TDHF has no parameter adjusted on reaction mechanism.
Some perspectives for extensions beyond TDHF to include cluster-transfers are discussed.
\end{abstract}
\maketitle

\label{sec:intro}

Binary collisions of many-body systems are of fundamental interest 
to test dynamical approaches of the quantum many-body problem. 
During the collision, the systems may retain their entities [(in)elastic scattering] or new ones may be produced
if they fuse or transfer some 
constituents.
Examples of transfer reactions include electron transfer in ion or cluster collisions~\cite{cam00}, 
and nucleon transfer in collisions of atomic nuclei~\cite{cor09}.
The prediction of the outcome of such reactions 
is one of the main challenges of modern quantum many-body dynamics theories.
In particular, the transfer products may be in a coherent superposition of fragments with different constituent numbers,
and 
transfer probabilities should be computed to allow comparison with experiments.


The coupled channel framework, where the relative motion of the collision partners is coupled to their internal degrees of freedom,
is 
amongst the most popular approaches to study transfer reactions~\cite{oer01,tho09}. 
It 
allows a detailed reproduction of experimental data, 
providing the fact that the structure of the collision partners 
(ground and excited states) as well as their interaction potential are well known.
For numerical tractability, however, only few states are usually included.
In addition, all information on the structure of the reactants is not always available, as, e.g., for exotic nuclei.
It is then important to develop other approaches with less parameters, to enhance their predictive power.
Recent works have pushed the envelope of describing binary collisions of many-body systems 
both quantum mechanically and microscopically, with no parameter adjusted on reaction mechanisms.
For instance, the dynamics of the valence electrons in collisions of atoms, molecules, or atomic clusters, 
is usually given by the time-dependent density functional theory (TDDFT) (see, e.g.,~\cite{wan09} and references therein).
In nuclear physics, these approaches usually consider independent particles evolving in a mean-field as a starting point, 
as in the time-dependent Hartree-Fock (TDHF) theory~\cite{kim97,uma06}.
Although they have been mostly applied to fusion reactions,
several recent attempts of describing nucleon transfer in heavy ion collisions within TDHF 
have been made~\cite{uma08a,sim08a,gol09,was09b,sim10,ked10}.


Here, we use a particle number projection technique on the fragments of the many-body state 
to determine the transfer probabilities. 
This technique is standard in beyond-mean-field models for nuclear structure 
when the number of particles is only given in average~\cite{rin80}.
In the present work, it is applied in the context of heavy-ion collisions,
however, it could be generalized to determine the particle number distribution 
in fragments of any many-body system, for instance, following electron transfer or ionization in atomic clusters, nuclear fission...

We investigate sequential transfer of nucleons in $^{16}$O+$^{208}$Pb collisions using the TDHF theory.
Nucleon transfer may occur when the projectile has enough energy 
to overcome the Coulomb repulsion and reach the vicinity of its collision partner, 
that is, at energies around and down to few MeV 
below the so-called fusion barrier.
Here, we focus on sub-barrier central collisions and compare our 
calculations 
with the sum of experimental one 
and two-proton transfer probabilities. 
Note that the relative yield between 
one and two-proton transfer is 
sensitive to nucleon clusters 
which are not included in TDHF.
Perspectives of this work in terms of beyond-TDHF improvements 
to treat properly correlations responsible for transfer of nucleon clusters are then discussed. 



The 
TDHF theory has been introduced by Dirac~\cite{dir30}.
In nuclear physics, it is usually used with a Skyrme energy density functional (EDF)~\cite{sky56}
to generate the nuclear mean-field~\cite{bon76,neg82,sim10}. 
 The EDF is the only phenomenological ingredient
which is adjusted on few nuclear structure properties~\cite{cha98}. 
The same EDF is used to compute the initial Hartree-Fock ground state 
of the nuclei and the time evolution.
The 
$N$ particles 
are constrained
to be in an anti-symmetrized independent particle state (Slater determinant) at any time.
The state vector reads $\kfi=\prod_{i=1}^N\oad_i|-\>$ where $\oad_i$ creates a particle in the state $|i\>$
when applied on the particle vacuum~$|-\>$.
The one-body density matrix of such a state reads 
$\rho(\mathbf{r} sq, \mathbf{r'}s'q') = \sum_{i} n_i \varphi_i^{sq}(\mathbf{r}){\varphi_i^{s'q'}}^*(\mathbf{r'})$, 
where $\varphi_i^{sq}(\mathbf{r})=\<{\vr}sq|i\>$ is a single-particle wave function, 
$\mathbf{r}$, $s$ and $q$ denote the nucleon position, spin, and isospin, respectively, 
and $n_i=1$ for occupied states 
($1\le{i}\le{N}$) and 0 otherwise.
The TDHF equation reads $i\hbar \frac{\partial}{\partial t} \rho = \left[h[\rho],\rho\right]$.
The single particle Hamiltonian $h[\rho]$ is related to the Skyrme EDF, noted $E[\rho]$, 
which depends on local densities~\cite{eng75} by 
 $h[\rho](\mathbf{r}sq, \mathbf{r'}s'q') = \frac{\delta E[\rho]}{\delta \rho(\mathbf{r'}s'q', \mathbf{r} sq)}$. 

Realistic TDHF calculations in 3 dimensions are now possible 
with modern Skyrme functionals including spin-orbit term~\cite{kim97,nak05,uma06,mar06}.  
Here, the TDHF equation is solved iteratively in time using the {\textsc{tdhf3d}} code 
with the SLy4$d$ parameterization of the Skyrme EDF~\cite{kim97}. 
This code is a time-dependent extension of a version of the {\textsc{ev8}} code without pairing~\cite{bon05}.
The algorithm for the time-evolution is described in~\cite{bon76,sim10}.
A time step $\Delta{t}=1.5\times10^{-24}$~s is used.
The spatial grid has $N_x\times{N_y}\times{N_z/2}=84\times28\times14$ points 
with a plane of symmetry (the collision plane $z=0$) and a lattice spacing $\Delta{x}=0.8$~fm.
The initial distance between the nuclei is~44.8~fm. 



\begin{figure}
\includegraphics[width=5cm]{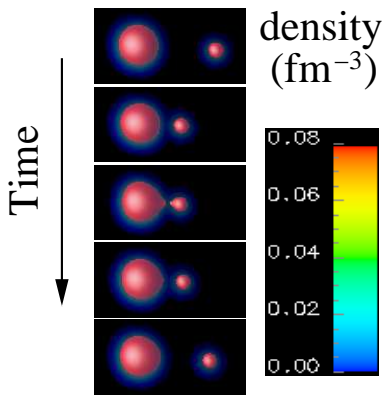} 
\caption{(color online) Density evolution for the central collision of a $^{16}$O (initially on the right side)
with a $^{208}$Pb (left) at $E_{c.m.}=74.44$~MeV.
The snapshots run from $t=7.5$ to 37.5~zs by steps of 7.5~zs.}
\label{fig:dens}
\end{figure}

The density evolution of the $^{16}$O+$^{208}$Pb central collision 
at a center of mass energy $E_{c.m.}=74.44$~MeV (just below the fusion barrier) plotted in Fig.~\ref{fig:dens} shows that
 the two nuclei form a di-nuclear system with a 
neck and then re-separate. There is {\it a priori} no reason that these two fragments conserve the
same average neutron and proton numbers as in the entrance channel~\cite{uma08a} (except for symmetric reactions).
Indeed, between the touching and re-separation, nucleons can be exchanged.
In TDHF calculations, this exchange is treated through the time-dependent distortion of single-particle
wave-functions which can eventually be partially transferred from one partner to the other.

The following operator written in r-space counts the number of particles with isospin $q$ in the right side of the separation plane   
(defined arbitrarily as $x>0$):
\oeq
\oN_R^q = \sum_{s} \sdf \int \stb \d \vr \stf \oad(\vr s q) \sdf \oa(\vr s q) 
\sdf \Theta(x)
\label{eq:NG}
\ceq
where $\Theta(x)=1$ if $x>0$ and 0 elsewhere, and
$\oa({\vr}sq)=\sum_i\az_i^{sq}(\vr) \oa_i$. 
Let us write 
$
\< i | j \>_R^q = \sum_{s} \int \sdb \d \vr \stf {\az_i^{sq}}^*(\vr) \sdf {\az_j^{sq}}(\vr)\sdf \Theta(x)
$
the overlap in the $x>0$ region between two single-particle states with isospin $q$. 
Using $\<\oad_i\oa_j\>=n_i\del_{ij}$, 
we obtain the average number of particles in the $x>0$ region as
$\<\oN_R^q\> 
= \sum_i \sdf \<i|i\>_R^q \sdf n_i.
$
Applied to the average proton and neutron numbers of the small fragment 
after a central collision at $E_{c.m.}=74.44$~MeV (see Fig. \ref{fig:dens}), 
we get $\sim6.1$~protons and $\sim8.1$~neutrons, respectively.
This indicates that the proton transfer probability from the light to the heavy fragment is so high at the barrier 
that two protons, in average, have been sequentially transferred.
Decreasing the energy induces a rapid convergence of the average proton and neutron numbers towards the $^{16}$O ones.
Indeed, $\<\oN_R^p\>\simeq\<\oN_R^n\>\simeq 8.0$ at $E_{c.m.}=70$~MeV.




Well below the barrier, where transfer is prohibited,  
the variance of $\oN_R$ is strictly zero:
$\si_R^2=\<\oN_R^2\>-\<\oN_R\>^2=0$
(here and in the following, we omit the isospin $q$ for simplicity).
This property is lost at higher energies where transfer occurs.
Then, the system in the exit channel is not an eigenstate of $\oN_R$, 
and each fragment is no longer described by an eigenstate of the particle number operator
(e.g., a Slater determinant).
Note that the upper limit of the variance obeys 
$\si_R^2  \le{\<\oN_R\>  \(1-\frac{\<\oN_R\>}{N_t}\)}$
for a Slater determinant~\cite{das79}, where $N_t$ is the total number of protons or neutrons.
This is an intrinsic limitation of independent particle systems.
In case of violent collisions such as deep-inelastic reactions, 
experimental variances may exceed this limit~\cite{das79}, and
inclusion of correlations is then needed.
However, for less violent collisions such as sub-barrier transfer reactions, 
smaller experimental variances are expected, 
and a mean-field approach like TDHF might give reasonable estimates of the variances.

Let us calculate the variance $\si_R^2$ after the reaction.
Using anti-commutation relations for fermions and $\<\oad_i\oad_j\oa_k\oa_l\>=n_in_j(\del_{il}\del_{jk}-\del_{ik}\del_{jl})$ for a Slater determinant, we get~\cite{das79}
$
\si_R^2 = \<\oN_R\> - \sum_{i,j=1}^N \sdf \lll\< i|j\>_R \rll^2.
$
Applying this formula to the small fragment in the exit channel
 of the reaction at $E_{c.m.}=74.44$~MeV shown in Fig.~\ref{fig:dens},
we get $\si_{R}^p\simeq0.5$ for protons and  $\si_{R}^n\simeq0.3$ for neutrons. 
At $E_{c.m.}=70$~MeV, we get $\si_{R}^p\simeq\si_{R}^n\simeq0.2$, showing that 
transfer occurs at this energy, although it does not change the average number of protons and neutrons as discussed before. 
These finite values of $\si_{R}$ clearly indicate that the many-body systems on each side of the 
separation plane are no longer eigenstates of the particle number operator. 


To get a deeper insight into these TDHF predictions, 
we now compute
the transfer probabilities.
It is possible to extract the component of the wave function associated to a specific transfer channel 
using a particle number projector onto $N$ protons or neutrons in the $x>0$ region.
Such a projector is written
$
\oP_R(N)=\frac{1}{2\pi}\int_0^{2\pi} \stb \d \tet \stf e^{i\tet(\oN_R-N)}
$~\cite{ben03}.
It can be used to compute the probability to find $N$ nucleons in $x>0$ in the state $\kfi$,
\oeq
\lll\oP_R(N)\kfi\rll^2=\frac{1}{2\pi}\int_0^{2\pi} \stb \d \tet \stf e^{-i\tet{N}}\bfi\phi_R(\tet)\>,
\label{eq:proba}
\ceq
where $|\phi_R(\tet)\>=e^{{i\tet\oN_R}}\kfi$.
Note that $|\phi_R(\tet)\>$ is an independent particle state. 
The last term in Eq.~(\ref{eq:proba}) is then the determinant of the matrix of the occupied single particle state overlaps:
$\bfi\phi_R(\tet)\>=\det (F)$
with
$$
F_{ij}= \sum_{s} \int \stb\d \vr \sdf{\az_i^s}^*(\vr) {\az_j^{s}}(\vr) e^{i\tet\Theta(x)}
=\del_{ij}+\<i|j\>_R(e^{i\tet}-1).
$$

\begin{figure}
\includegraphics[width=7cm]{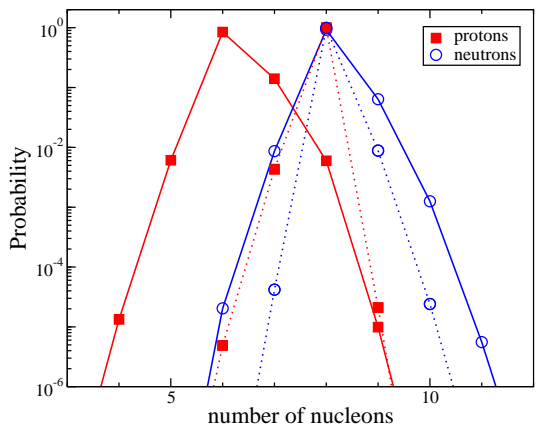} 
\caption{Neutron (circles) and proton (squares) number probability distributions 
of the lightest fragment in exit channel of a head-on 
$^{16}$O+$^{208}$Pb collision at $E_{c.m.}=74.44$~MeV (solid lines) and 65~MeV (dotted lines). }
\label{fig:barrier}
\end{figure}

The integral in Eq.~(\ref{eq:proba}) is discretized using $\tet_n=2\pi{n}/M$ with the integer $n=1\cdots{M}$.
Choosing $M=300$ ensures convergence. 
The resulting probabilities are shown in Fig.~\ref{fig:barrier}
for central collisions at $E_{c.m.}=74.44$~MeV and 65~MeV ($\sim13\%$ below the barrier).
At the barrier, the most probable channel is a two-proton transfer leading to a $^{14}$C nucleus in the exit channel.
At the lower energy, the transfer probabilities are typically one or several orders of magnitude lower than at the barrier
and the (in)elastic channels are by far the dominant ones. 
Note that the probability for proton stripping (transfer from the light to the heavy nucleus) 
is higher than for proton pickup (transfer from the heavy to the light nucleus) as observed experimentally~\cite{vid77},
while neutron pickup is more probable than neutron stripping.


In transfer experiments, one usually measures
angular differential cross-sections 
for 
multi-nucleon transfer channels.
It is numerically heavy and time consuming to compute such cross-sections.
A standard alternative is to translate the experimental angular cross-sections at sub-barrier energies into transfer probabilities 
as a function of the distance of closest approach $R_{min}$ 
between the collision partners assuming a Rutherford trajectory~\cite{bro91}:
$
R_{min}={Z_1Z_2e^2}[1+\mbox{cosec}(\al/2)]/{2E_{c.m.}}
$
where $\al$ is the center of mass scattering angle, and $Z_{1,2}$ the proton numbers of the colliding nuclei.
Experimental transfer probabilities can then be calculated from the ratio of sub-barrier transfer 
to Rutherford cross-sections~\cite{bro91} for a given 
 distance of closest approach.

\begin{figure}
\includegraphics[width=7cm]{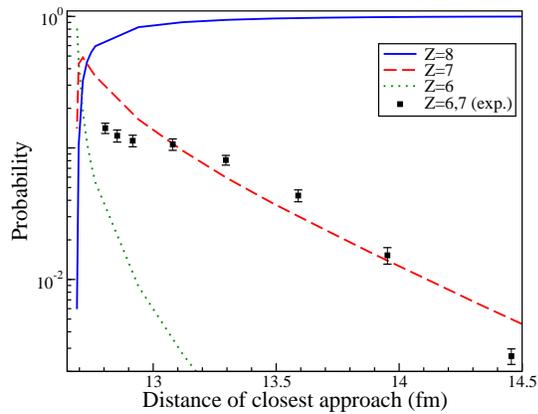} 
\caption{Proton number probability distribution as function of the distance of closest approach
obtained with TDHF (lines). Experimental data (squares) are adapted from Ref.~\cite{vid77}
using $E_{c.m.}	=74.3$~MeV data and show the sum of the one and two-proton transfer channels.}
\label{fig:p_e}
\end{figure}

The evolutions of the main proton-transfer channels with the distance of closest approach
predicted by TDHF for head-on collisions are shown in Fig.~\ref{fig:p_e} in solid, dashed and dotted lines 
for zero, one and two-proton stripping, respectively.
In fact, the TDHF probability for two-proton transfer behaves roughly 
as the square of the one-proton transfer probability (if the latter is small compared to one), 
which is a signature for sequential transfer~\cite{cor09}.
The two-proton transfer in TDHF is, then, much smaller than the one-proton one
(except at the barrier, corresponding to $R_{min}\simeq12.7$~fm).

Multi-proton transfer has been measured for $^{16}$O+$^{208}$Pb at $E_{c.m.}=74.3$~MeV by Videb{\ae}k {\it et~al.}~\cite{vid77}.
One and two-proton stripping has been observed at this energy, and no proton-pickup, 
in qualitative agreement with TDHF calculations.
However, it is well known that the two-proton stripping in this reaction 
occurs mainly as a cluster transfer, i.e., as a pair or alpha-transfer~\cite{tho89,vid77}.
The treatment of such nucleon-clusters involves correlations beyond TDHF.
As a consequence, TDHF is not expected to reproduce the ratio between the one and two-proton transfer probabilities,
but only their sums which should be less affected by such cluster structures.
Indeed, in a simple model with transfer probability per nucleon $p\ll1$ and $xN$ (resp. $(1-x)N$) paired (unpaired) nucleons, 
one expects the two-nucleon transfer probability to be $P_{2n}\sim{xNp}$ 
(as the two correlated nucleons are transferred as a cluster) and the one-nucleon transfer to be $P_{1n}\sim{(1-x)Np}$.
The sum $P_{1n}+P_{2n}\sim{Np}$ is then independent on the correlations.  
The experimental sum of the one and two-proton transfer probabilities are shown in Fig.~\ref{fig:p_e} (squares).
They can be compared with the one-proton transfer in TDHF, as the two-proton sequential transfer is negligible in this energy range. 
The overall agreement is good, considering the fact that 
TDHF has no parameter adjusted on reaction mechanism. 
Note that the overestimation of the data at $R_{min}<13$~fm might be due to the fact that sub-barrier fusion is not included in TDHF while it would remove some flux from quasi-elastic channels at distances close to the barrier radius.

Pair transfer should be enhanced by pairing correlations and
could be investigated with the TDHF-Bogolyubov theory~\cite{ave08,eba10}, 
or the time-dependent density-matrix theory~\cite{ass09}.
Variances of fragment mass and charge distributions would be improved with
stochastic techniques to account for zero point motion~\cite{was09b}
or the Balian-V\'en\'eroni variational approach~\cite{bro08}.
One limitation of TDHF is that all exit channels follow the same trajectory. 
Several TDHF trajectories with different external potentials ''forcing'' transfer 
could be used to build a more general state using the time-dependent Generator Coordinate Method~\cite{gou05}. 
In addition, dynamical eikonal approximation could be used to account
of quantal interferences between different trajectories~\cite{bay05}.
Finally, investigations of the excitation energies of the transfer products should be studied 
with, e.g., the density-constrained TDHF approach~\cite{uma09} (see also~\cite{was09}). 

\begin{acknowledgments}
The calculations have been performed on the Centre de Calcul Recherche et Technologie of the Commissariat \`a l'\'Energie Atomique, France. 
M.~Dasgupta, M.~Evers, D.~J.~Hinde, D. Lacroix, and B. Avez are thanked for discussions 
 and a careful reading of the paper. 
\end{acknowledgments}

\bibliography{transfer}

\begin{thebibliography}{36}
\expandafter\ifx\csname natexlab\endcsname\relax\def\natexlab#1{#1}\fi
\expandafter\ifx\csname bibnamefont\endcsname\relax
  \def\bibnamefont#1{#1}\fi
\expandafter\ifx\csname bibfnamefont\endcsname\relax
  \def\bibfnamefont#1{#1}\fi
\expandafter\ifx\csname citenamefont\endcsname\relax
  \def\citenamefont#1{#1}\fi
\expandafter\ifx\csname url\endcsname\relax
  \def\url#1{\texttt{#1}}\fi
\expandafter\ifx\csname urlprefix\endcsname\relax\def\urlprefix{URL }\fi
\providecommand{\bibinfo}[2]{#2}
\providecommand{\eprint}[2][]{\url{#2}}

\bibitem[{\citenamefont{Campbell and Rohmund}(2000)}]{cam00}
\bibinfo{author}{\bibfnamefont{E.~E.~B.} \bibnamefont{Campbell}}
  \bibnamefont{and} \bibinfo{author}{\bibfnamefont{F.}~\bibnamefont{Rohmund}},
  \bibinfo{journal}{Rep. Prog. Phys.} \textbf{\bibinfo{volume}{63}},
  \bibinfo{pages}{1061} (\bibinfo{year}{2000}).

\bibitem[{\citenamefont{Corradi et~al.}(2009)\citenamefont{Corradi, Pollarolo,
  and Szilner}}]{cor09}
\bibinfo{author}{\bibfnamefont{L.}~\bibnamefont{Corradi}},
  \bibinfo{author}{\bibfnamefont{G.}~\bibnamefont{Pollarolo}},
  \bibnamefont{and} \bibinfo{author}{\bibfnamefont{S.}~\bibnamefont{Szilner}},
  \bibinfo{journal}{J. Phys. G} \textbf{\bibinfo{volume}{36}},
  \bibinfo{pages}{113101} (\bibinfo{year}{2009}).

\bibitem[{\citenamefont{von Oertzen and Vitturi}(2001)}]{oer01}
\bibinfo{author}{\bibfnamefont{W.}~\bibnamefont{von Oertzen}} \bibnamefont{and}
  \bibinfo{author}{\bibfnamefont{A.}~\bibnamefont{Vitturi}},
  \bibinfo{journal}{Rep. Prog. Phys.} \textbf{\bibinfo{volume}{64}},
  \bibinfo{pages}{1247} (\bibinfo{year}{2001}).

\bibitem[{\citenamefont{Thompson and Nunes}(2009)}]{tho09}
\bibinfo{author}{\bibfnamefont{I.~J.} \bibnamefont{Thompson}} \bibnamefont{and}
  \bibinfo{author}{\bibfnamefont{F.~M.} \bibnamefont{Nunes}},
  \emph{\bibinfo{title}{Nuclear Reactions for Astrophysics: Principles,
  Calculation and Applications of Low-Energy Reactions}}
  (\bibinfo{publisher}{Cambridge University Press}, \bibinfo{year}{2009}), ISBN
  \bibinfo{isbn}{0521856353}.

\bibitem[{\citenamefont{Wang et~al.}(2009)\citenamefont{Wang, Dinh, Reinhard,
  Suraud, Bruny, Montano, Feil, Eden, Abdoul-Carime, Farizon et~al.}}]{wan09}
\bibinfo{author}{\bibfnamefont{Z.}~\bibnamefont{Wang}},
  \bibinfo{author}{\bibfnamefont{P.}~\bibnamefont{Dinh}},
  \bibinfo{author}{\bibfnamefont{P.-G.} \bibnamefont{Reinhard}},
  \bibinfo{author}{\bibfnamefont{E.}~\bibnamefont{Suraud}},
  \bibinfo{author}{\bibfnamefont{G.}~\bibnamefont{Bruny}},
  \bibinfo{author}{\bibfnamefont{C.}~\bibnamefont{Montano}},
  \bibinfo{author}{\bibfnamefont{S.}~\bibnamefont{Feil}},
  \bibinfo{author}{\bibfnamefont{S.}~\bibnamefont{Eden}},
  \bibinfo{author}{\bibfnamefont{H.}~\bibnamefont{Abdoul-Carime}},
  \bibinfo{author}{\bibfnamefont{B.}~\bibnamefont{Farizon}},
  \bibnamefont{et~al.}, \bibinfo{journal}{Int. J. Mass Spectrom.}
  \textbf{\bibinfo{volume}{285}}, \bibinfo{pages}{143 } (\bibinfo{year}{2009}).

\bibitem[{\citenamefont{Kim et~al.}(1997)\citenamefont{Kim, Otsuka, and
  Bonche}}]{kim97}
\bibinfo{author}{\bibfnamefont{K.-H.} \bibnamefont{Kim}},
  \bibinfo{author}{\bibfnamefont{T.}~\bibnamefont{Otsuka}}, \bibnamefont{and}
  \bibinfo{author}{\bibfnamefont{P.}~\bibnamefont{Bonche}},
  \bibinfo{journal}{J. Phys. G} \textbf{\bibinfo{volume}{23}},
  \bibinfo{pages}{1267} (\bibinfo{year}{1997}).

\bibitem[{\citenamefont{Umar and Oberacker}(2006)}]{uma06}
\bibinfo{author}{\bibfnamefont{A.~S.} \bibnamefont{Umar}} \bibnamefont{and}
  \bibinfo{author}{\bibfnamefont{V.~E.} \bibnamefont{Oberacker}},
  \bibinfo{journal}{Phys. Rev. C} \textbf{\bibinfo{volume}{73}},
  \bibinfo{pages}{054607} (\bibinfo{year}{2006}).

\bibitem[{\citenamefont{Umar et~al.}(2008)\citenamefont{Umar, Oberacker, and
  Maruhn}}]{uma08a}
\bibinfo{author}{\bibfnamefont{A.}~\bibnamefont{Umar}},
  \bibinfo{author}{\bibfnamefont{V.}~\bibnamefont{Oberacker}},
  \bibnamefont{and} \bibinfo{author}{\bibfnamefont{J.}~\bibnamefont{Maruhn}},
  \bibinfo{journal}{Eur. Phys J. A} \textbf{\bibinfo{volume}{37}},
  \bibinfo{pages}{245} (\bibinfo{year}{2008}).

\bibitem[{\citenamefont{Simenel and Avez}(2008)}]{sim08a}
\bibinfo{author}{\bibfnamefont{C.}~\bibnamefont{Simenel}} \bibnamefont{and}
  \bibinfo{author}{\bibfnamefont{B.}~\bibnamefont{Avez}},
  \bibinfo{journal}{Int. J. Mod. Phys. E} \textbf{\bibinfo{volume}{17}},
  \bibinfo{pages}{31} (\bibinfo{year}{2008}).

\bibitem[{\citenamefont{Golabek and Simenel}(2009)}]{gol09}
\bibinfo{author}{\bibfnamefont{C.}~\bibnamefont{Golabek}} \bibnamefont{and}
  \bibinfo{author}{\bibfnamefont{C.}~\bibnamefont{Simenel}},
  \bibinfo{journal}{Phys. Rev. Lett.} \textbf{\bibinfo{volume}{103}},
  \bibinfo{pages}{042701} (\bibinfo{year}{2009}).

\bibitem[{\citenamefont{Washiyama
  et~al.}(2009{\natexlab{a}})\citenamefont{Washiyama, Ayik, and
  Lacroix}}]{was09b}
\bibinfo{author}{\bibfnamefont{K.}~\bibnamefont{Washiyama}},
  \bibinfo{author}{\bibfnamefont{S.}~\bibnamefont{Ayik}}, \bibnamefont{and}
  \bibinfo{author}{\bibfnamefont{D.}~\bibnamefont{Lacroix}},
  \bibinfo{journal}{Phys. Rev. C} \textbf{\bibinfo{volume}{80}},
  \bibinfo{pages}{031602} (\bibinfo{year}{2009}{\natexlab{a}}).

\bibitem[{\citenamefont{Simenel et~al.}(2010)\citenamefont{Simenel, Lacroix,
  and Avez}}]{sim10}
\bibinfo{author}{\bibfnamefont{C.}~\bibnamefont{Simenel}},
  \bibinfo{author}{\bibfnamefont{D.}~\bibnamefont{Lacroix}}, \bibnamefont{and}
  \bibinfo{author}{\bibfnamefont{B.}~\bibnamefont{Avez}},
  \emph{\bibinfo{title}{Quantum Many-Body Dynamics: Applications to Nuclear
  Reactions}} (\bibinfo{publisher}{VDM Verlad}, \bibinfo{year}{2010}).

\bibitem[{\citenamefont{Kedziora and Simenel}(2010)}]{ked10}
\bibinfo{author}{\bibfnamefont{D.~J.} \bibnamefont{Kedziora}} \bibnamefont{and}
  \bibinfo{author}{\bibfnamefont{C.}~\bibnamefont{Simenel}},
  \bibinfo{journal}{Phys. Rev. C} \textbf{\bibinfo{volume}{81}},
  \bibinfo{pages}{044613} (\bibinfo{year}{2010}).

\bibitem[{\citenamefont{Ring and Schuck}(1980)}]{rin80}
\bibinfo{author}{\bibfnamefont{P.}~\bibnamefont{Ring}} \bibnamefont{and}
  \bibinfo{author}{\bibfnamefont{P.}~\bibnamefont{Schuck}},
  \emph{\bibinfo{title}{The Nuclear Many-Body Problem (Theoretical and
  Mathematical Physics)}} (\bibinfo{publisher}{Springer},
  \bibinfo{year}{1980}), ISBN \bibinfo{isbn}{354021206X}.

\bibitem[{\citenamefont{Dirac}(1930)}]{dir30}
\bibinfo{author}{\bibfnamefont{P.~A.~M.} \bibnamefont{Dirac}},
  \bibinfo{journal}{Proc. Camb. Phil. Soc.} \textbf{\bibinfo{volume}{26}},
  \bibinfo{pages}{376} (\bibinfo{year}{1930}).

\bibitem[{\citenamefont{Skyrme}(1956)}]{sky56}
\bibinfo{author}{\bibfnamefont{T.}~\bibnamefont{Skyrme}},
  \bibinfo{journal}{Phil. Mag.} \textbf{\bibinfo{volume}{1}},
  \bibinfo{pages}{1043} (\bibinfo{year}{1956}).

\bibitem[{\citenamefont{Bonche et~al.}(1976)\citenamefont{Bonche, Koonin, and
  Negele}}]{bon76}
\bibinfo{author}{\bibfnamefont{P.}~\bibnamefont{Bonche}},
  \bibinfo{author}{\bibfnamefont{S.}~\bibnamefont{Koonin}}, \bibnamefont{and}
  \bibinfo{author}{\bibfnamefont{J.~W.} \bibnamefont{Negele}},
  \bibinfo{journal}{Phys. Rev. C} \textbf{\bibinfo{volume}{13}},
  \bibinfo{pages}{1226} (\bibinfo{year}{1976}).

\bibitem[{\citenamefont{Negele}(1982)}]{neg82}
\bibinfo{author}{\bibfnamefont{J.~W.} \bibnamefont{Negele}},
  \bibinfo{journal}{Rev. Mod. Phys.} \textbf{\bibinfo{volume}{54}},
  \bibinfo{pages}{913} (\bibinfo{year}{1982}).

\bibitem[{\citenamefont{Chabanat et~al.}(1998)\citenamefont{Chabanat, Bonche,
  Haensel, Meyer, and Schaeffer}}]{cha98}
\bibinfo{author}{\bibfnamefont{E.}~\bibnamefont{Chabanat}},
  \bibinfo{author}{\bibfnamefont{P.}~\bibnamefont{Bonche}},
  \bibinfo{author}{\bibfnamefont{P.}~\bibnamefont{Haensel}},
  \bibinfo{author}{\bibfnamefont{J.}~\bibnamefont{Meyer}}, \bibnamefont{and}
  \bibinfo{author}{\bibfnamefont{R.}~\bibnamefont{Schaeffer}},
  \bibinfo{journal}{Nuclear Physics A} \textbf{\bibinfo{volume}{635}},
  \bibinfo{pages}{231 } (\bibinfo{year}{1998}).

\bibitem[{\citenamefont{Engel et~al.}(1975)\citenamefont{Engel, Brink, Goeke,
  Krieger, and Vautherin}}]{eng75}
\bibinfo{author}{\bibfnamefont{Y.~M.} \bibnamefont{Engel}},
  \bibinfo{author}{\bibfnamefont{D.~M.} \bibnamefont{Brink}},
  \bibinfo{author}{\bibfnamefont{K.}~\bibnamefont{Goeke}},
  \bibinfo{author}{\bibfnamefont{S.~J.} \bibnamefont{Krieger}},
  \bibnamefont{and}
  \bibinfo{author}{\bibfnamefont{D.}~\bibnamefont{Vautherin}},
  \bibinfo{journal}{Nuc. Phys. A} \textbf{\bibinfo{volume}{249}},
  \bibinfo{pages}{215 } (\bibinfo{year}{1975}).

\bibitem[{\citenamefont{Nakatsukasa and Yabana}(2005)}]{nak05}
\bibinfo{author}{\bibfnamefont{T.}~\bibnamefont{Nakatsukasa}} \bibnamefont{and}
  \bibinfo{author}{\bibfnamefont{K.}~\bibnamefont{Yabana}},
  \bibinfo{journal}{Phys. Rev. C} \textbf{\bibinfo{volume}{71}},
  \bibinfo{pages}{024301} (\bibinfo{year}{2005}).

\bibitem[{\citenamefont{Maruhn et~al.}(2006)\citenamefont{Maruhn, Reinhard,
  Stevenson, and Strayer}}]{mar06}
\bibinfo{author}{\bibfnamefont{J.~A.} \bibnamefont{Maruhn}},
  \bibinfo{author}{\bibfnamefont{P.-G.} \bibnamefont{Reinhard}},
  \bibinfo{author}{\bibfnamefont{P.~D.} \bibnamefont{Stevenson}},
  \bibnamefont{and} \bibinfo{author}{\bibfnamefont{M.~R.}
  \bibnamefont{Strayer}}, \bibinfo{journal}{Phys. Rev. C}
  \textbf{\bibinfo{volume}{74}}, \bibinfo{pages}{027601}
  (\bibinfo{year}{2006}).

\bibitem[{\citenamefont{Bonche et~al.}(2005)\citenamefont{Bonche, Flocard, and
  Heenen}}]{bon05}
\bibinfo{author}{\bibfnamefont{P.}~\bibnamefont{Bonche}},
  \bibinfo{author}{\bibfnamefont{H.}~\bibnamefont{Flocard}}, \bibnamefont{and}
  \bibinfo{author}{\bibfnamefont{P.}~\bibnamefont{Heenen}},
  \bibinfo{journal}{Comp. Phys. Comm.} \textbf{\bibinfo{volume}{171}},
  \bibinfo{pages}{49 } (\bibinfo{year}{2005}).

\bibitem[{\citenamefont{Dasso et~al.}(1979)\citenamefont{Dasso, D{\o}ssing, and
  Pauli}}]{das79}
\bibinfo{author}{\bibfnamefont{C.~H.} \bibnamefont{Dasso}},
  \bibinfo{author}{\bibfnamefont{T.}~\bibnamefont{D{\o}ssing}},
  \bibnamefont{and} \bibinfo{author}{\bibfnamefont{H.~C.} \bibnamefont{Pauli}},
  \bibinfo{journal}{Z. Phys. A} \textbf{\bibinfo{volume}{289}},
  \bibinfo{pages}{395} (\bibinfo{year}{1979}).

\bibitem[{\citenamefont{Bender et~al.}(2003)\citenamefont{Bender, Heenen, and
  Reinhard}}]{ben03}
\bibinfo{author}{\bibfnamefont{M.}~\bibnamefont{Bender}},
  \bibinfo{author}{\bibfnamefont{P.-H.} \bibnamefont{Heenen}},
  \bibnamefont{and} \bibinfo{author}{\bibfnamefont{P.-G.}
  \bibnamefont{Reinhard}}, \bibinfo{journal}{Rev. Mod. Phys.}
  \textbf{\bibinfo{volume}{75}}, \bibinfo{pages}{121} (\bibinfo{year}{2003}).

\bibitem[{\citenamefont{Videb\ae{}k et~al.}(1977)\citenamefont{Videb\ae{}k,
  Goldstein, Grodzins, Steadman, Belote, and Garrett}}]{vid77}
\bibinfo{author}{\bibfnamefont{F.}~\bibnamefont{Videb\ae{}k}},
  \bibinfo{author}{\bibfnamefont{R.~B.} \bibnamefont{Goldstein}},
  \bibinfo{author}{\bibfnamefont{L.}~\bibnamefont{Grodzins}},
  \bibinfo{author}{\bibfnamefont{S.~G.} \bibnamefont{Steadman}},
  \bibinfo{author}{\bibfnamefont{T.~A.} \bibnamefont{Belote}},
  \bibnamefont{and} \bibinfo{author}{\bibfnamefont{J.~D.}
  \bibnamefont{Garrett}}, \bibinfo{journal}{Phys. Rev. C}
  \textbf{\bibinfo{volume}{15}}, \bibinfo{pages}{954} (\bibinfo{year}{1977}).

\bibitem[{\citenamefont{Broglia and Winther}(1991)}]{bro91}
\bibinfo{author}{\bibfnamefont{R.~A.} \bibnamefont{Broglia}} \bibnamefont{and}
  \bibinfo{author}{\bibfnamefont{A.}~\bibnamefont{Winther}},
  \emph{\bibinfo{title}{Heavy Ion Reactions: Lecture Notes the Elementary
  Processes (Frontiers in Physics)}} (\bibinfo{publisher}{Addison Wesley
  Publishing Company, New-York}, \bibinfo{year}{1991}), ISBN
  \bibinfo{isbn}{0201513927}.

\bibitem[{\citenamefont{Thompson et~al.}(1989)\citenamefont{Thompson,
  Nagarajan, Lilley, and Smithson}}]{tho89}
\bibinfo{author}{\bibfnamefont{I.~J.} \bibnamefont{Thompson}},
  \bibinfo{author}{\bibfnamefont{M.~A.} \bibnamefont{Nagarajan}},
  \bibinfo{author}{\bibfnamefont{J.~S.} \bibnamefont{Lilley}},
  \bibnamefont{and} \bibinfo{author}{\bibfnamefont{M.~J.}
  \bibnamefont{Smithson}}, \bibinfo{journal}{Nucl. Phys. A}
  \textbf{\bibinfo{volume}{505}}, \bibinfo{pages}{84 } (\bibinfo{year}{1989}).

\bibitem[{\citenamefont{Avez et~al.}(2008)\citenamefont{Avez, Simenel, and
  Chomaz}}]{ave08}
\bibinfo{author}{\bibfnamefont{B.}~\bibnamefont{Avez}},
  \bibinfo{author}{\bibfnamefont{C.}~\bibnamefont{Simenel}}, \bibnamefont{and}
  \bibinfo{author}{\bibfnamefont{P.}~\bibnamefont{Chomaz}},
  \bibinfo{journal}{Phys. Rev. C} \textbf{\bibinfo{volume}{78}},
  \bibinfo{pages}{044318} (\bibinfo{year}{2008}).

\bibitem[{\citenamefont{Ebata et~al.}(2010)\citenamefont{Ebata, Nakatsukasa,
  Inakura, Yoshida, Hashimoto, and Yabana}}]{eba10}
\bibinfo{author}{\bibfnamefont{S.}~\bibnamefont{Ebata}},
  \bibinfo{author}{\bibfnamefont{T.}~\bibnamefont{Nakatsukasa}},
  \bibinfo{author}{\bibfnamefont{T.}~\bibnamefont{Inakura}},
  \bibinfo{author}{\bibfnamefont{K.}~\bibnamefont{Yoshida}},
  \bibinfo{author}{\bibfnamefont{Y.}~\bibnamefont{Hashimoto}},
  \bibnamefont{and} \bibinfo{author}{\bibfnamefont{K.}~\bibnamefont{Yabana}},
  \bibinfo{journal}{Phys. Rev. C} \textbf{\bibinfo{volume}{82}},
  \bibinfo{pages}{034306} (\bibinfo{year}{2010}).

\bibitem[{\citenamefont{Assi\'e and Lacroix}(2009)}]{ass09}
\bibinfo{author}{\bibfnamefont{M.}~\bibnamefont{Assi\'e}} \bibnamefont{and}
  \bibinfo{author}{\bibfnamefont{D.}~\bibnamefont{Lacroix}},
  \bibinfo{journal}{Phys. Rev. Lett.} \textbf{\bibinfo{volume}{102}},
  \bibinfo{pages}{202501} (\bibinfo{year}{2009}).

\bibitem[{\citenamefont{Broomfield and Stevenson}(2008)}]{bro08}
\bibinfo{author}{\bibfnamefont{J.~M.~A.} \bibnamefont{Broomfield}}
  \bibnamefont{and} \bibinfo{author}{\bibfnamefont{P.~D.}
  \bibnamefont{Stevenson}}, \bibinfo{journal}{J. Phys. G}
  \textbf{\bibinfo{volume}{35}}, \bibinfo{pages}{095102}
  (\bibinfo{year}{2008}).

\bibitem[{\citenamefont{Goutte et~al.}(2005)\citenamefont{Goutte, Berger,
  Casoli, and Gogny}}]{gou05}
\bibinfo{author}{\bibfnamefont{H.}~\bibnamefont{Goutte}},
  \bibinfo{author}{\bibfnamefont{J.~F.} \bibnamefont{Berger}},
  \bibinfo{author}{\bibfnamefont{P.}~\bibnamefont{Casoli}}, \bibnamefont{and}
  \bibinfo{author}{\bibfnamefont{D.}~\bibnamefont{Gogny}},
  \bibinfo{journal}{Phys. Rev. C} \textbf{\bibinfo{volume}{71}},
  \bibinfo{pages}{024316} (\bibinfo{year}{2005}).

\bibitem[{\citenamefont{Baye et~al.}(2005)\citenamefont{Baye, Capel, and
  Goldstein}}]{bay05}
\bibinfo{author}{\bibfnamefont{D.}~\bibnamefont{Baye}},
  \bibinfo{author}{\bibfnamefont{P.}~\bibnamefont{Capel}}, \bibnamefont{and}
  \bibinfo{author}{\bibfnamefont{G.}~\bibnamefont{Goldstein}},
  \bibinfo{journal}{Phys. Rev. Lett.} \textbf{\bibinfo{volume}{95}},
  \bibinfo{pages}{082502} (\bibinfo{year}{2005}).

\bibitem[{\citenamefont{Umar et~al.}(2009)\citenamefont{Umar, Oberacker,
  Maruhn, and Reinhard}}]{uma09}
\bibinfo{author}{\bibfnamefont{A.~S.} \bibnamefont{Umar}},
  \bibinfo{author}{\bibfnamefont{V.~E.} \bibnamefont{Oberacker}},
  \bibinfo{author}{\bibfnamefont{J.~A.} \bibnamefont{Maruhn}},
  \bibnamefont{and} \bibinfo{author}{\bibfnamefont{P.~G.}
  \bibnamefont{Reinhard}}, \bibinfo{journal}{Phys. Rev. C}
  \textbf{\bibinfo{volume}{80}}, \bibinfo{pages}{041601}
  (\bibinfo{year}{2009}).

\bibitem[{\citenamefont{Washiyama
  et~al.}(2009{\natexlab{b}})\citenamefont{Washiyama, Lacroix, and
  Ayik}}]{was09}
\bibinfo{author}{\bibfnamefont{K.}~\bibnamefont{Washiyama}},
  \bibinfo{author}{\bibfnamefont{D.}~\bibnamefont{Lacroix}}, \bibnamefont{and}
  \bibinfo{author}{\bibfnamefont{S.}~\bibnamefont{Ayik}},
  \bibinfo{journal}{Phys. Rev. C} \textbf{\bibinfo{volume}{79}},
  \bibinfo{pages}{024609} (\bibinfo{year}{2009}{\natexlab{b}}).

\end{thebibliography}

\end{document}